\documentstyle[12pt]{article}

\def\lsim{\raise0.3ex\hbox{$<$\kern-0.75em\raise-1.1ex\hbox{$\sim$}}}
\def\gsim{\raise0.3ex\hbox{$>$\kern-0.75em\raise-1.1ex\hbox{$\sim$}}}
\def\noi{\noindent}
\def\to{\rightarrow}

\def\ie{{\it i.e.}}

\def\beq{\begin{equation}} \def\eeq{\end{equation}}
\def\bea{\begin{eqnarray}} \def\eea{\end{eqnarray}}
\def\nn{\nonumber}
\def\noi{\noindent}
\def\beeq{\begin{eqnarray}} \def\eeeq{\end{eqnarray}}
\newcommand\mysection{\setcounter{equation}{0}\section}

\newcounter{hran} 
\def\what{\widehat}
\def\vev#1{\langle #1 \rangle}

\def\gev{~{\rm GeV}}
\def\fbi{~{\rm fb}^{-1}}

\def\anti{\overline}
\def\tanb{\tan\beta} 
\def\br{BR}
\def\gam{\gamma}
\def\nsd#1{$N_{SD}(#1)$}

\begin{document}

\begin{center}
{\Large\bf Establishing a No-Lose Theorem for \\ [.21in] NMSSM 
Higgs Boson Discovery at the LHC}
\end{center}
\vskip 0.5 truecm

\centerline{\bf Ulrich Ellwanger}\vskip 1 truemm
\centerline{Laboratoire de Physique Th\'eorique\footnote{Unit\'e Mixte
de Recherche - CNRS - UMR 8627}} 
\centerline{Universit\'e de Paris XI, B\^atiment 210, F-91405 ORSAY
Cedex, France}
\vskip 2 truemm
\centerline{\bf John F. Gunion}
\vskip 1 truemm
\centerline{Davis Institute for High Energy Physics} 
\centerline{University of California, Davis, California 95616, USA}
\vskip 2 truemm
\centerline{\bf Cyril Hugonie}
\vskip 1 truemm
\centerline{Institute for Particle Physics Phenomenology} 
\centerline{University of Durham, DH1 3LE, UK}

\begin{abstract}
  
  We scan the parameter space of the NMSSM for the observability of at
  least one Higgs boson at the LHC with $300\fbi$ integrated
  luminosity, taking the present LEP2 constraints into account. We
  restrict the scan to those regions of parameter space for which
  Higgs boson decays to other Higgs bosons and/or supersymmetric
  particles are kinematically forbidden.  We find that if $WW$-fusion
  detection modes for a light Higgs boson are not taken into account,
  then there are still significant regions in the scanned portion of
  the NMSSM parameter space where no Higgs boson can be observed at
  the $5\sigma$ level, despite the recent improvements in ATLAS and
  CMS procedures and techniques and even if we combine all non-fusion
  discovery channels. However, if the $WW$-fusion detection modes are
  included using the current theoretical study estimates, then we find
  that for all scanned points at least one of the NMSSM Higgs bosons
  will be detected. If the estimated $300\fbi$ significances for ATLAS
  and CMS are combined, one can also achieve $5\sigma$ signals after
  combining just the non-$WW$-fusion channels signals. We present the
  parameters of several particularly difficult points, and discuss the
  complementary roles played by different modes. We conclude that the
  LHC will discover at least one NMSSM Higgs boson unless there are
  large branching ratios for decays to SUSY particles and/or to other
  Higgs bosons.

\end{abstract}

\vskip 1. truecm
\noi LPT Orsay 01-100 \par
\noi UCD-01-13 \par
\noi RAL-TR-2001-035 \par 
\noi IPPP/01/45 \par 
\noi hep-ph/0111179 \par
\noi October, 2001 \par

\newpage
\pagestyle{plain}
\baselineskip 18pt

\mysection{Introduction}

Supersymmetric extensions of the standard model generally predict relatively
light Higgs bosons. One of the most important tasks of the LHC is the search
for Higgs bosons \cite{1r,2r}. An important milestone in understanding the
potential of the LHC was the demonstration that at least one Higgs boson of
the minimal supersymmetric standard model (MSSM) would be detectable at the
$\geq 5\sigma$ level throughout all of the MSSM parameter space so long as top
squark masses do not exceed 1.5 to 2 TeV and so long as large branching
fractions to decay channels containing supersymmetric particles are not
substantial.

In the present paper, we study, subject to these same and a few other
simplifying restrictions, the detectability of Higgs bosons in the
next-to-minimal supersymmetric standard model (NMSSM). In the NMSSM, one Higgs
singlet superfield, $\what S$, is added to the MSSM in order to render
unnecessary the bilinear superpotential term $\mu \what H_1 \what H_2$ by
replacing it with $\lambda \what S\what H_1 \what H_2 $, where the vacuum
expectation value of the scalar component of $\what S$, $\vev{S}$, results in
an effective bilinear Higgs mixing with $\mu=\lambda\vev{S}$. The detectability
of the NMSSM Higgs bosons was first considered in a contribution to Snowmass 96
\cite{3r}. The result, using the experimentally established modes and
sensitivities available at the time, was that substantial regions in the
parameter space of the NMSSM were found where none of the Higgs bosons would
have been observable either at LEP2 or at the LHC even with an integrated
luminosity of $600\fbi$ (two detectors with $L=300\fbi$ each). 

Since then, progress has been made both on the theoretical and the experimental
sides. On the theoretical side, the dominant two-loop corrections to the
effective potential of the model have been computed \cite{4r,5r}. These lead to
a modest decrease in the mass of the lightest Higgs scalar, holding fixed the
stop sector parameters. Inclusion of the two-loop corrections thus increases
somewhat the part of the NMSSM parameter space excluded by LEP2 (and accessible
at the Tevatron) \cite{5r}, but is of less relevance for the LHC. On the
experimental side the expected statistical significances have been improved
since 1996 \cite{1r,2r}. Most notably, associated $t {\anti t} h$ production
with $h \to b{\anti b}$ (originally discussed in \cite{6r}), which in the SM
context is particularly sensitive to $m_h\ \lsim\ 120$ GeV, has been added by
ATLAS and CMS to the list of Higgs boson detection modes
\cite{1r,2r,6.01r,6.02r}. Analysis of this mode was recently extended
\cite{6.1r} to $m_h=140\gev$, which, though not relevant in the SM case due to
the decline in the $b\anti b$ branching ratio as the $WW^*$ mode increases, is
highly relevant for points in our searches for which the $WW^*$ mode is
suppressed in comparison to the SM prediction. In addition, techniques have
been proposed \cite{6.2r} for isolating signals for $WW$ fusion to a light
Higgs boson which decays to $\tau\anti\tau$ or $WW^{(*)}$.

It turns out that adding in just the $t\anti t h$ process renders the
no-Higgs-discovery parameter choices described and plotted in \cite{3r},
including the ``black point'' described in detail there, visible \cite{7r}. In
the present paper, we search for any remaining parameter choices for which no
Higgs boson would produce a $\geq 5\sigma$ signal. In this search, we perform a
scan over nearly all of the parameter space of the model, the only parameter
choices not included being those for which there is sensitivity to the highly
model-dependent decays of Higgs bosons to other Higgs bosons and/or
superparticles. The outcome is that, for an integrated luminosity of $300\fbi$
at the LHC, there are still regions in the parameter space with $<5 \sigma$
expected statistical significance (computed as $N_{SD}=S/\sqrt B$ for a given
mode) for all Higgs detection modes so far studied in detail by ATLAS and CMS,
\ie\ including the $t\anti t h\to t\anti t b\anti b$ mode but not the
$WW$-fusion modes. On the other hand, the expected statistical significance for
at least one of these detection modes is always above $3.6 \sigma$ at
$300\fbi$, and the statistical significance obtained by combining (using the
naive Gaussian procedure) all the non-$WW$-fusion modes is at least $4.8
\sigma$. However, we find that all such cases are quite observable (at $\geq
10.1 \sigma$) in one of the $WW$-fusion modes (using theoretically estimated
statistical significances for these modes). For all points in the scan of
parameter space, statistical significances obtained by combining all modes,
including $WW$-fusion modes, are always $\gsim 10.7 \sigma$. Thus, NMSSM Higgs
discovery by just one detector with $L=300\fbi$ is essentially guaranteed for
those portions of parameter space for which Higgs decays to other Higgs bosons
or supersymmetric particles are kinematically forbidden. This represents
substantial progress towards guaranteeing LHC discovery of at least one of the
NMSSM Higgs bosons.

In order to clarify the nature of the most difficult points in those portions
of parameter space considered, we present, in sect. 4, examples of particularly
difficult bench mark points for the Higgs sector of the NMSSM. Apart from the
``bare'' parameters of the model, we give the masses and couplings of all Higgs
scalars, their production rates and branching ratios to various channels
(relative to the SM Higgs) and details of the statistical significances
predicted for each Higgs boson in each channel. The latter will allow an
assessment of exactly what level of improvement in statistical significance
will be required in the various different detection modes in order to render
marginal modes visible. Of course, our estimates of the expected statistical
significances are often somewhat crude (e.g. their dependence on the
accumulated integrated luminosity). We believe that our procedures always err
in the conservative direction, leading to statistical significances that might
be a bit small. Thus, the LHC procedures for isolating Higgs boson signals
could provide even more robust signals for NMSSM Higgs boson detection than we
estimate here.

The detection modes, which serve for the searches for standard model or MSSM
Higgs bosons, include (using the notation $h$, $a$ for CP-even, CP-odd Higgs
bosons, respectively):\par
\noi 1) $gg \to h \to \gamma \gamma$;\par
\noi 2) associated $W h$ or $t \bar{t}h$ production with $\gamma \gamma
\ell^{\pm}$ in the final state;\par
\noi 3) associated $t {\anti t}h$ production with $h \to b{\anti b}$;\par
\noi 4) $gg \to h/a$ or associated $b {\anti b}h/a$ production with $h/a \to
\tau {\anti \tau}$;\par
\noi 5) $gg \to h \to ZZ^{(*)} \to$ 4 leptons;\par
\noi 6) $gg \to h \to WW^{(*)} \to l^+ l^- \nu {\anti \nu}$;\par
\noi 7) LEP2 $e^+e^- \to Zh$ and $e^+e^- \to ha$;\par
\noi 8) $WW\to h \to \tau\anti\tau$;\par
\noi 9) $WW\to h\to WW^{(*)}$,\par
\noi where 8) and 9) are those analyzed at the theoretical level in \cite{6.2r}
and included in the NMSSM analysis for the first time in this paper. The above
detection modes do not employ the possibly important decay channels i)~$h \to
hh$, ii)~$h \to aa$, iii)~$h \to h^+h^-$, iv)~$h \to aZ$, v)~$a \to ha$, vi)~$a
\to hZ$, vii)~$h,a \to h^\pm W^\mp$, viii)~$h,a \to t\anti t$ and ix)~$t \to
h^+b$. The decay modes i)-vii) give high multiplicity final states and
deserve a dedicated study \cite{wip}, while the existing analyses of the
$t\anti t$ final state signatures are not very detailed. Further, when
kinematically allowed, the $t\to h^+b$ signal would be easily observed
according to existing analyzes. Thus, in this paper we restrict our scan over
NMSSM parameter space to those parameters for which none of these decays are
present. In addition, we take the constraints of LEP2 [via the mode 7)] into
account, and only accept points for which $5\sigma$ discovery at LEP2 would not
have been possible \cite{10r,10.1r}.

The Higgs sector of the NMSSM consists of 3 scalars, denoted $h_1, h_2, h_3$
with $m_{h_1} < m_{h_2} < m_{h_3}$, 2 pseudo-scalars, denoted $a_1, a_2$ with
$m_{a_1} < m_{a_2}$, and a charged Higgs pair, denoted $h^{\pm}$. Mixing of the
neutral doublet fields with the gauge singlet fields in the scalar and in the
pseudo-scalar sector can be strong. The scalar mixing can lead to a
simultaneous suppression of the couplings of all the $h_i$ to gauge bosons, and
hence to a suppression of many of the detection modes above. (Of course, the
$a_i$ have no tree-level couplings to gauge boson pairs and the one-loop
couplings are too small to yield useful event rates.) The couplings of the
Higgs bosons to t- or b-quarks can be amplified, reduced or even change sign
with respect to the standard model couplings. Hence negative interferences can
occur among the (loop-) diagrams contributing to $gg \to h_i$ and $h_i \to
\gamma \gamma$, leading again to suppressions of the above detection modes. A
complete simultaneous annulation of all detection modes is not possible, but
simultaneous reduction of all detection modes is possible and it is for such
parameter choices that NMSSM Higgs boson discovery is most difficult.

In the next section, we define the class of models we are going to consider,
and the way we perform the scan over the corresponding parameter space. In
section 3 we describe our computations of the expected statistical
significances of the detection modes 1) -- 9) above. In section 4, we present 
six particularly difficult bench mark points (in table 1) and details
regarding their statistical significances in channels 1)-9) in table 2, with a
summary of overall statistical significances in table 3. Using these tables, we
give a discussion of the properties of these points.

\mysection{NMSSM Parameters and Scanning Procedure}

In this paper, we consider the simplest version of the NMSSM
\cite{7.1r,8r,9r}, where the term $\mu \what H_1 \what H_2$ in the
superpotential of the MSSM is replaced by (we use the notation $\what A$ for
the superfield and $A$ for its scalar component field)
\beq
\label{2.1r}
\lambda \what H_1 \what H_2 \what S\ + \ {\kappa \over 3} \what S^3 \ \ ,
\eeq
so that the superpotential is scale invariant. We make no assumption on
``universal'' soft terms. Hence, the five soft supersymmetry breaking terms
\beq
\label{2.2r}
m_{H_1}^2 H_1^2\ +\ m_{H_2}^2 H_2^2\ +\ m_S^2 S^2\ +\ \lambda
A_{\lambda}H_1 H_2 S\ +\ {\kappa \over 3} A_{\kappa}S^3
\eeq
are considered as independent. The masses and/or couplings of sparticles are
assumed to be such that their contributions to the loop diagrams inducing Higgs
production by gluon fusion and Higgs decay into $\gamma \gamma$ are negligible.
In the stop sector, which appears in the radiative corrections to the Higgs
potential, we chose the soft masses $m_Q = m_T \equiv M_{susy}= 1$ TeV, and
varied the stop mixing parameter 
\beq
\label{2.4r}
X_t \equiv 2 \ \frac{A_t^2}{M_{susy}^2+m_t^2} \left ( 1 -
\frac{A_t^2}{12(M_{susy}^2+m_t^2)} \right ) \ .
\eeq 
As in the MSSM, the value $X_t = \sqrt{6}$ -- so called maximal mixing --
maximizes the radiative corrections to the Higgs masses, and we found that it
leads to the most challenging points in the parameter space of the NMSSM. 

Assuming that the Higgs sector is CP conserving, the independent parameters of 
the model are thus: $\lambda, \kappa, m_{H_1}^2, m_{H_2}^2, m_S^2, A_{\lambda}$
and $A_{\kappa}$. For purposes of scanning and analysis, it is more convenient
to eliminate $m_{H_1}^2$, $m_{H_2}^2$ and $m_S^2$ in favor of $M_Z$,
$\tan\beta$ and $\mu_{\rm eff} = \lambda \langle S \rangle$ through the three
minimization equations of the Higgs potential (including the dominant 1- and
2-loop corrections \cite{5r}) and to scan over the six independent parameters 
\beq
\label{2.5r}
\lambda, \kappa, \tan\beta, \mu_{\rm eff}, A_{\lambda}, A_{\kappa}\ .
\eeq
We adopt the convention $\lambda,\kappa>0$, in which $\tanb$ can have either
sign. The absence of Landau singularities for $\lambda$ and $\kappa$ below the
GUT scale ($\sim 2\times10^{16}$ GeV) imposes upper bounds on these couplings
at the weak scale, which depend on the value of $h_t$ and hence of $\tanb$
\cite{7.1r,8r}. Using $m_{top}^{pole} = 175$ GeV, one finds $\lambda_{\rm
max}\sim 0.69$ and $\kappa_{\rm max}\sim 0.62$ for intermediate values of
$\tanb$.

For each point in the parameter space, we diagonalize the scalar and
pseudo-scalar mass matrices and compute the scalar, pseudo-scalar and charged
Higgs masses and couplings taking into account the dominant 1- and 2-loop
radiative corrections \cite{5r}. We then demand that the Higgs scalars satisfy
the LEP2 constraints on the $e^+e^- \to Zh_i$ production mode (taken from
\cite{10r}, fig. 10), which gives a lower bound on $m_{h_i}$ as a function of
the $ZZh_i$ reduced coupling. We also impose LEP2 constraints on $e^+e^- \to
h_ia_j$ associated production (from \cite{10.1r}, fig. 6), yielding a lower
bound on $m_{h_i} + m_{a_j}$ as a function of the $Zh_ia_j$ reduced coupling.

In order to render the above-mentioned processes i) -- ix) kinematically
impossible, we require the following inequalities among the masses:
\bea
& m_{h_3} < 2m_{h_1}, \ 2m_{a_1}, \ 2m_{h^\pm}, \ m_{a_1}+M_Z, \ m_{h^\pm}+M_W;
\nn \\
& m_{a_2} < m_{h_1}+m_{a_1}, \ m_{h_1}+M_Z, \ m_{h^\pm}+M_W; \quad m_{h^\pm} >
155 \mbox{GeV}. \nn
\eea
\noi In addition we require $|\mu_{\rm eff}|\ >\ 100$ GeV; otherwise a light
chargino would have been detected at LEP2. (The precise lower bound on
$|\mu_{\rm eff}|$ depends somewhat on $\tan\beta$ and the precise experimental
lower bound on the chargino mass; however, our subsequent results do not depend
on the precise choice of the lower bound on $|\mu_{\rm eff}|$.) We further note
that for the most challenging parameter space points that we shall shortly
discuss, $|\mu_{\rm eff}|\ >\ 100\gev$ is already sufficient to guarantee that
the NMSSM Higgs bosons cannot decay to chargino pairs so long as the SU(2)
soft-SUSY-breaking parameter $M_2$ is also large. In fact, in order to avoid
significant corrections to $\gam\gam h_i$ and $\gam\gam a_i$ couplings coming
from chargino loops it is easiest to take $M_2\gg \mu_{\rm eff}$ (or vice
versa). This is because the $h_i \widetilde \chi^+_i \widetilde \chi^-_i$
coupling is suppressed if the $\widetilde \chi^+_i$ is either pure higgsino or
pure gaugino. Since the parts of parameter space that are challenging with
regard to Higgs detection typically have $|\mu|\sim 100 - 200\gev$, the
validity of our assumptions requires that $M_2$ be large and that the chargino
be essentially pure higgsino.

Using a very rough sampling, we determined, as expected from previous work
\cite{3r}, that it is only for moderate values of $\tanb$ that $<5\sigma$
signals might possibly occur. From this sampling, we determined the most
difficult parameter space regions and further refined our scan to the
following:
\begin{itemize}
\itemsep=0in
\item
$4.5<|\tanb|<8$ (both signs) in steps of 0.25;
\item
$0.001<\lambda<{\rm min}[0.21,\lambda_{\rm max}]$, using 20 points;
\item
$0.001<\kappa<{\rm min}[0.24,\kappa_{\rm max}]$, using 20 points; 
\item
$100\gev<|\mu_{\rm eff}|<300\gev$ (both signs), in steps of 5 GeV;
\item
$0<|A_\lambda|<160\gev$, with $A_\lambda$ opposite in sign to $\mu_{\rm eff}$,
using steps of 5 GeV;
\item
$25\gev<|A_\kappa|<170\gev$, with $A_\kappa$ opposite in sign to $\mu_{\rm
eff}$, using steps of 5 GeV.
\end{itemize}
For those points sampled in this final scan which satisfy all the constraints
detailed earlier, we compute the expected statistical significances for the
processes 1) to 9) listed in section 1, as described in the next section. As a
rough guide, from the $\sim 10^9$ points detailed in the above list, we find
about $250,000$ that pass all constraints and have $N_{SD}<5$ (for $L=300\fbi$)
in each of the individual discovery modes 1) -- 7). We shall tabulate a number
of representative points taken from this final set in section 4.

\mysection{Expected Statistical Significances}

From the known couplings of the NMSSM Higgs scalars to gauge bosons and
fermions it is straightforward to compute their production rates in gluon-gluon
fusion and various associated production processes, as well as their partial
widths into $\gamma \gamma$, gauge bosons and fermions, either relative to a
standard model Higgs scalar or relative to the MSSM $H$ and/or $A$. This allows
us to apply ``NMSSM corrections'' to the processes 1) -- 9) above.

These NMSSM corrections are computed in terms of the following ratios. For the
scalar Higgs bosons, $c_V$ is the ratio of the coupling of the $h_i$ to vector
bosons as compared to that of a SM Higgs boson (the coupling ratios for $h_iZZ$
and $h_iWW$ are the same), and $c_t$, $c_b$ are the corresponding ratios of
the couplings to top and bottom quarks (one has $c_\tau=c_b$). Note that we
always have $|c_V| < 1$, but $c_t$ and $c_b$ can be larger, smaller or even
differ in sign with respect to the standard model. For the CP-odd Higgs bosons,
$c_V$ is not relevant since there is no tree-level coupling of the $a_i$ to the
$VV$ states; $c_t$ and $c_b$ are defined as the ratio of the $i\gamma_5$
couplings for $t\anti t$ and $b\anti b$, respectively, relative to SM-like
strength.

We emphasize that our procedure implicitely includes QCD
corrections to the Higgs production processes at precisely the same
level as the experimental collaborations.
First, the ATLAS and CMS collaborations employed Monte Carlo programs
such as ISAJET \cite{ISAJET} and PYTHIA \cite{PYTHIA} in obtaining
results for the (MS)SM. These programs include many QCD corrections to
Higgs production in a leading-log sense. This is the best that can
currently be done to implement QCD corrections in the context of
experimental cuts and neural-net analyses. Clearly the more exact NNLO
results for many of the relevant processes will slowly be implemented
in the Monte Carlo programs and increased precision for Higgs discovery
expectations will result. Since our goal is to obtain NMSSM results
that are completely analogous to the currently available (MS)SM
results, we have proceeded by simply rescaling the available (MS)SM
experimental analyses. 
%MODIFIED
In doing the rescaling of the Higgs branching ratios we have included
all relevant higher-order QCD corrections \cite{radcor} using an
adapted version of the FORTRAN code HDECAY \cite{HDECAY}. 
\footnote{In our
  computations however, we neglect the contribution to the
  $h_i\gam\gam$ coupling coming from the charged Higgs loop. Despite
  the relatively small masses of the $h^+$ for our most problematical
  points, the charged Higgs loop decouples \cite{11.1r}, especially
  for the small values of $\lambda$ and $\lambda A_{\lambda}$
  characteristic of difficult points for which the actual $h^+h^- h_i$
  coupling is only of order a few times $g m_W/(4\sqrt 2)$
  \cite{7.1r}. Its contribution would typically only be of order a few
  percent even though our difficult points have smaller $\gam\gam h_i$
  coupling than a SM-like Higgs boson by virtue of suppressed $h_i WW$
  coupling and/or cancellation between the top and $W$ loop
  contributions.}  We now give additional details on our rescaling
procedures.

The expected statistical significances for the processes 1) and 6) are computed
beginning with results for the SM Higgs boson taken from ref \cite{11r}, fig.1
(``Expected Observability of Standard Model Higgs in CMS with $100\fbi\,$'').
The application of the NMSSM corrections using $c_V$, $c_t$ and $c_b$ [which
determine $\Gamma(gg\to h_i)$, $\br(h_i\to\gam\gam)$ and $\br(h_i\to WW^*)$]
is straightforward in these two cases.

The expected statistical significances for process 2) are taken from the same
figure. In ref. \cite{12r} one finds that $W h_i$ and $t \bar{t}h_i$ production
contribute with roughly equal weight to the SM signal. This allows us to
decompose the expected significance into the corresponding production
processes, apply the NMSSM corrections, and then recombine the production
processes.

The expected Standard Model Higgs statistical significances for process 3) are
taken from table 19-8 in ref. \cite{1r}, with the extension to Higgs masses
above 120 GeV as provided in \cite{6.1r}, using a numerical interpolation for
Higgs masses below 140 GeV. For the standard model process 5) we again use ref.
\cite{1r}, tables 19-18 and 19-21. In both cases, the application of the NMSSM
corrections is straightforward.

The estimation of the statistical significances for the process 4) in the NMSSM
requires the most discussion. Figure 19-62 of ref. \cite{1r} and fig. 8 of ref.
\cite{12r} give the $5 \sigma$ contours in the $\tan\beta$ - $m_A$ plane of the
MSSM. The critical issue is how much of these $5\sigma$ signals derive from
$gg\to H+gg\to A$ production and how much from associated $b\anti b H+b\anti
bA$ production, and how each of the $gg$ fusion and $b\anti b$ associated
production processes are divided up between $H$ and $A$. For the former, we
turn to table 19-35 of ref. \cite{1r}. There, we see that it is for cuts
designed to single out the associated production processes that large
statistical significance can be achieved and that such cuts provide 90\% of the
net statistical significance of $N_{SD}=8.9$ (3.9 for $gg$ fusion cuts combined
in quadrature with 8.0 for $b\anti bH+b\anti bA$ associated production cuts)
for $m_A=150\gev$ and $L=30\fbi$. (For the associated production cuts, the
table shows that the contribution of the $gg$ fusion processes to the signal is
very small.) The percentage of $N_{SD}$ deriving from $gg$-fusion cuts is even
smaller at high $m_A$. Since we are mainly interested in $m_H,\ m_A\in
[100\gev,200\gev]$, we will assume that 90\% of the statistical significance
along the contours of fig. 19-62 comes from the associated production cut
analysis; this will give us a slightly conservative estimate of the associated
production $N_{SD}$ values at still higher $m_A$. With this choice, the
$5\sigma$ contour at $L=100\fbi$ from fig. 19-62 of ref. \cite{1r} corresponds
to a $4.5\sigma$ contour for associated $b\anti b H+b\anti bA$ production
alone. Since the values of $\tanb$ along this contour are large, we can
separate the $H$ and $A$ signals from one another by using the following
properties of the MSSM within which fig. 19-62 of ref. \cite{1r} was
generated: (a) $\br(H\to \tau\anti\tau)\sim \br(A\to\tau\anti\tau)\sim 0.09$; 
(b) the $b\anti b A$ and $b\anti b H$ couplings are very nearly equal and scale
as $\tan\beta$; and (c) $m_A\sim m_H$ within the $\tau\anti\tau$ mass
resolution. As a result, the net signal rate along this contour is
approximately twice that for $b\anti b A$ or $b\anti b H$ alone. Thus,
$N_{SD}=2.25$ would be achieved for $b\anti b A$ or $b\anti b H$ along this
contour were $m_A$ and $m_H$ widely separated.

We can then compute the statistical significance for the $b\anti b h_i$ and
$b\anti b a_i$ signals with $h_i,a_i\to \tau\anti\tau$ decay using the
following procedure. First, the NMSSM $b\anti b h_i$ and $b\anti b a_i$
production rates are related to the MSSM $b\anti b H$ and $b\anti b A$ rates 
by the factors $[c_b(h_i)]^2 / \tan^2\beta$ and $[c_b(a_i)]^2 / \tan^2\beta$,
respectively. Next, we account for the fact that the $\tau\anti \tau$
branching ratios of the NMSSM scalars and pseudo-scalars differ somewhat from
the value of $0.09$ appropriate for the MSSM $H$ and $A$. In particular,
$\br(h_3\to \tau\anti\tau)$ is significantly reduced when the $h_3$ has large
enough mass and large enough $c_V$ that it acquires a modest $WW^*$ branching
ratio. Typical reductions will be tabulated in table 2. Thus, defining the
value of $\tanb$ as a function of $m_A$ shown by the $100\fbi$ curve of fig.
19-62 in ref. \cite{1r} as $\tanb_{2.25}(m_A)$, we compute $N_{SD}({h_i})$ for
$L=100\fbi$ as
\beq
N_{SD}(h_i)=2.25 \left[{c_b(h_i)\over \tanb_{2.25}(m_{h_i})}\right]^2
\br_{\tau\anti\tau}(h_i)\,, 
\eeq
where $\br_{\tau\anti \tau}\equiv \br(h_i\to \tau\anti\tau)/\br(H\to\tau\anti
\tau) =\br(h_i\to \tau\anti\tau)/0.09$. An exactly parallel procedure is
employed for $N_{SD}(a_i)$.

The above procedure is conservative in that it assumes no contribution to the
$\tau\anti\tau$ channel $N_{SD}$ from the $gg$ fusion processes. However, as we
shall describe in the next section, for the most difficult points in parameter
space the $gg$-fusion rates are very substantially suppressed relative to MSSM
values. For these points, essentially 99\% of the $\tau\anti\tau$ channel
$N_{SD}$ would derive from $b\anti b$+Higgs associated production.

In recombining the scalar and pseudo-scalar signals, we must account for the
fact that they can have fairly different masses in the NMSSM. In this paper, we
have chosen to recombine the scalar and pseudo-scalar signals at different
masses following the procedure of ref. \cite{13r}, section 5.4, with $\sigma_m
\sim 30$ GeV as estimated from fig. 19-61 in \cite{1r} at high luminosity and
extrapolated to $m_A\ \lsim\ 150$ GeV. This procedure leads to somewhat
approximate estimates of the NMSSM statistical significances for this
detection mode.

Results for the statistical significances 
of the $h_i$ signals in modes 8) and 9)
were similarly obtained by rescaling the theoretical results of 
\cite{6.2r} ( summarized most conveniently in
the last of the listed papers)
using the values of $[c_V(h_i)]^2$, $\br{\tau\anti\tau}$ and $\br{WW^*}$. 

Using the above procedures, for each point in the parameter space of the NMSSM
we obtain the statistical significances predicted for an integrated luminosity
of $100\fbi$ for each of the detection modes 1) -- 9). In order to obtain the
statistical significances for the various detection modes at $300\fbi$, we
multiply the $100\fbi$ statistical significances by $\sqrt{3}$ in the cases
1), 2), 3), 5) and 6), but only by a factor of $1.3$ in the cases 4), 8) and
9). That such a factor is appropriate for mode 4), see, for example, fig. 19-62
in \cite{1r}. Use of this same factor for modes 8) and 9) is simply a
conservative guess.

\mysection{Difficult Points}

As stated in the introduction we still find ``black spots'' in the parameter
space of the NMSSM, where the expected statistical significances for all Higgs
detection modes 1) -- 7) are below $5 \sigma$ at $300\fbi$. The reasons for
this phenomenon have been described above; see also the corresponding
discussion in \cite{3r}. However, after including the modes 8) and 9), the
points that provide the worst 1) -- 6) statistical significances typically
yield robust signals in one or the other of the $WW$-fusion modes 8) and 9).

In order to render the corresponding suppression mechanisms of the detection
modes reproducible, we present the detailed properties of several difficult
points in the parameter space in table 1. The notation is as follows: The bare
parameters are as in eq. (2.5), with $m_{H_1}^2$, $m_{H_2}^2$ and $m_S^2$ fixed
implicitly by the minimization conditions. (As noted earlier, with the
convention $\lambda$, $\kappa > 0$ in the NMSSM, the sign of $\tan\beta$ can no
longer be defined to be positive.) For the reasons discussed below eq.
(\ref{2.4r}) we chose in the stop sector $m_Q = m_T \equiv M_{susy}= 1$ TeV and
$X_t=\sqrt{6}$ for all of the points (1 -- 6). We have also fixed
$m_{top}^{pole}=175$ GeV. For both scalar and pseudoscalar Higgs bosons, ``gg
Production Rate'' denotes the ratio of the gluon-gluon production rate with
respect to that obtained if $c_t=c_b=1$, keeping the Higgs mass fixed. For
scalar $h_i$, this is the same as the ratio of the $gg$ production rate
relative to that predicted for a SM Higgs boson of the same mass. For the
scalar $h_i$, $\br\gamma \gamma$ denotes the ratio of the $\gamma \gamma$
branching ratio with respect to that of a SM Higgs boson with the same mass. (A
verification of the reduced gluon-gluon production rates or $\gamma \gamma$
branching ratios would sometimes require the knowledge of the couplings to
higher precision than given, for convenience, in table 1.) Also given for the
scalar $h_i$ are the ratios $\br b\anti b$ and $\br WW^*$ of the $b\anti b$ and
$WW^*$ branching ratios relative to the SM prediction (as noted above, one has
$\br\tau\anti\tau=\br b\anti b$). 

In table 2, we tabulate the statistical significances for the $h_i$ in all the
channels 1) -- 9); production of the CP-odd $a_i$ turns out to be relevant only
when they add to the $h_i$ signals in process 4). Also note that, all these
problematical points are such that $m_{h_1}+m_{a_1}>206\gev$, so that
$e^+e^-\to h_1+a_1$ followed by $h_1,a_1 \to b\anti b$ would have been
kinematically forbidden at the highest LEP2 energy. Hence, for LEP2 mode 7) we
only give the statistical significance for $e^+e^- \to Z h_i$. Also tabulated
in table 2 are four statistical significances obtained by combining various
channels. This combination is done in the Gaussian approximation:
\beq
N_{SD}^{\rm combined}= \left[\sum_{i} \left(N_{SD}^i\right)^2\right]^{1/2} \, ,
\nn
\eeq
where $\sum_i$ runs over the channels $i$ being combined. We give results for
the following combinations:
\begin{description}
\itemsep=-.05in
\item{a)} $N_{SD}$ obtained by combining LHC channels 1) -- 6); 
\item{b)} $N_{SD}$ obtained by combining LHC channels 1) -- 6) and LEP2; 
\item{c)} $N_{SD}$ obtained by combining LHC channels 1) -- 6) with the
$WW$-fusion channels 8) and 9);
\item{d)} $N_{SD}$ obtained by combining all LHC channels and LEP2, \ie\ by
combining all channels 1) -- 9).
\end{description}
In those cases where there is no LEP2 signal, a)=b) and c)=d). In addition, in
our point selection we have required a mass difference of at least 10 GeV
between scalar Higgses, so that they yield well separated signals and no
statistical significance combination of two different scalar Higgses is needed.
All parameter choices for which Higgs boson masses differ by less than
10 GeV yield stronger signals than the cases retained. 
(The increased net signal
strength of overlapping Higgs signals in those
channels with limited mass resolution arises as a result
of $N_{SD}^{\rm eff}(1+2)\sim (S_1+S_2)/\sqrt B>\sqrt{S_1^2+S_2^2}/\sqrt B$.) 

As summarized in table 3, all of the tabulated ``bench mark points'' have
statistical significances below $5 \sigma$ for all of the detection modes 1) --
6) at $300\fbi$ and 7) at LEP2. In more detail, as tabulated in table 2 and
summarized in table 3, the best signals in the modes 1) -- 6) for the points
\#1 -- \#6 at the LHC are:
\begin{itemize}
\itemsep=-.1in
\item
point \#1, $N_{SD}=$4.37 for mode 2) and $h_1$;
\item
point \#2, $N_{SD}=$3.95 for mode 3) and $h_2$;
\item
point \#3, $N_{SD}=$3.62 for mode 4) and $h_3$; 
\item
point \#4, $N_{SD}=$4.46 for mode 5) and $h_3$;
\item 
point \#5, $N_{SD}=$4.83 for mode 3) and $h_1$; 
\item
point \#6, $N_{SD}=$4.86 for mode 4) and $h_3$;
\end{itemize}
Further, for point \#3, the combined statistical significance of modes 1) -- 6)
(also tabulated in table 3) would still be below $5$ for any one $h_i$, 
although $\sqrt 2 N_{SD}^{1-6}>5$ (as is likely to be relevant by combining
ATLAS and CMS data once each detector has accumulated $L=300\fbi$) for at least
one of the $h_i$. However, for all these ``difficult'' points the $WW$-fusion
modes 8) and/or 9) provide (according to theoretical estimates) a decent
(sometimes very strong) signal.

The points \#1 -- \#4 differ as to which of the modes 1) -- 6) and which $h_i$
yields the largest statistical significance should the $WW$-fusion mode 8) not
provide as strong a signal as suggested by the theoretical estimates. To render
these points observable without the $WW$-fusion mode 8) would require
improvements of all detection modes 2) -- 5). 

As in \cite{3r}, we find that difficult points in the parameter space generally
have $|\tan\beta| \sim 5$. This is the region of $\tan\beta$ for which the
$b\anti b h,b\anti b a$ signals are still not very much enhanced but yet the 
$gg\to h,a$ and $t\anti t h,t\anti t a$ signals have been suppressed somewhat.
In a few cases, however, difficulties also arise for $|\tan\beta|$ as large as
8, as shown in the case of point \#5. Also as in \cite{3r}, the most difficult
points are those in which the masses of the $h_i$ and $a_i$ are relatively
close in magnitude, typically clustered in a $\sim 60\gev$ interval above $\sim
105\gev$. Such clustering maximizes the mixing among the different Higgs bosons
and thereby minimizes the significance of the discovery channels for any one
Higgs boson. In particular, it is for strong mixing among the $h_i$ that the
statistical significance for discovery modes based on a large $VV$ coupling for
any one $h_i$ are most easily suppressed.

Finally, for point \#6, we have minimized the statistical significances for the
$WW$-fusion modes over the parameter space, while keeping the statistical
significances of modes 1) -- 6) below 5. One can see that it still gives a
strong $10.1 \sigma$ signal in mode 8). 
[Smaller $N_{SD}$ for mode 8) would have been possible if we had allowed 
stronger signals in modes 1) -- 6), in particular had we allowed smaller
mass separation, $<10\gev$, between the two lightest Higgs bosons.]
In addition, for point \#6 $m_{h_1} =
112$ GeV and the $ZZ$ coupling of $h_1$ is sufficiently large that it would
have yielded a $4.8 \sigma$ signal at LEP2. Had we taken a top quark mass
slightly larger, $m_{top}^{pole}=178$ GeV, we would have found a very similar
point with a $h_1$ mass of $\sim 115$ GeV, which could have been responsible
for the excess observed at LEP2 \cite{14r}.

\mysection{Discussion and Conclusions}

In this paper, we have addressed the question of whether or not it would be
possible to fail to discover any of the Higgs bosons of the NMSSM using
combined LEP2 and LHC data, possibly resulting in the erroneous conclusion that
Higgs bosons with masses below 200 GeV have been excluded. We have demonstrated
that, assuming that the decay channels i) -- ix) are either kinematically
disallowed or render a Higgs boson observable, this is unlikely (at the
$>5\sigma$ level) to happen. Certainly, there are points in NMSSM parameter
space for which the statistical significances for the individual detection
modes 1) -- 6) (\ie\ those analyzed in detail by ATLAS and CMS) are all well
below $5 \sigma$ for integrated luminosity of $300\fbi$. However, by combining
several of the modes 1) -- 6) and $300\fbi$ data from both ATLAS and CMS, a
$>5\sigma$ signal can be achieved based just on modes 1) -- 6). Further, we
have found that throughout all of the NMSSM parameter space (scanned subject to
the earlier listed restrictions) for which such weak signals in modes 1) -- 6)
are predicted, the theoretical estimates for the $WW$-fusion modes indicate
that an easily detected $WW\to h\to \tau\anti \tau$ signal should be present.
Thus, our conclusion is that for all of the parameter space of the NMSSM
compatible with reasonable boundary conditions for the parameters at the GUT
scale (with, of course, non-universal soft terms in general) and such that
Higgs pair and SUSY pair decays of the Higgs bosons are kinematically
forbidden, at least one of the NMSSM Higgs bosons will be detected at the LHC.
This is a big improvement over the results from the earlier Snowmass 1996 study
which was somewhat negative without the inclusion of the $t\anti t h\to t\anti
t b\anti b$ mode 3), and the $WW$-fusion modes 8) and 9).

It is amusing to note that all of our bench mark points for which Higgs
discovery is most difficult at the LHC include a Higgs scalar with mass close
to 115 GeV (with, however, reduced couplings to the $Z$ boson), which could be
responsible for the excess observed at LEP2 \cite{14r}.

Another important point that appears from our analysis is the fact that the
full $L=300\fbi$ of integrated luminosity (per detector) is needed in order to
have robust NMSSM Higgs discovery in the portion of parameter space considered
here. Of course, as in the MSSM, it is very possible that only one of the
CP-even NMSSM Higgs bosons might be detected at the LHC but that, as studied by
Kamoshita et al. in \cite{9r}, the observation of all the CP-even Higgs bosons
of the NMSSM would be possible at the LC by virtue of all having some
non-negligible level of $ZZ$ coupling and not having very high masses. Even at
the LC, the CP-odd Higgs bosons might escape discovery, although this would not
be the case for the parameter choices that we have found which make LHC
discovery of even one NMSSM Higgs bosons most challenging. This is because, for
such parameters, the $a_i$ are relatively light and could be readily seen at
the LC in the processes $e^+e^-\to h_i a_j$, $e^+e^-\to \nu\anti \nu a_ia_i$
and $e^+e^-\to Z^*\to Z a_ia_i$, assuming an integrated LC luminosity of
$1000\fbi$ and energy $\sqrt s\geq 500\gev$ \cite{15r}.

This study makes clear the importance of continuing to expand the sensitivity
of existing modes and continuing to develop new modes for Higgs detection at
the LHC in order not to have to wait for construction of a linear $e^+e^-$
collider for detection of at least one of the SUSY Higgs bosons. In particular,
study of modes i) -- ix) and SUSY pair channels should all be pushed. The
problematical points that we have emphasized here are unlikely to be
substantially influenced by $t\anti t $ or SUSY decays since all the Higgs
masses are below $\sim 200\gev$ so that $t\anti t$ decays will be kinematically
highly suppressed (one of the top quarks would have to be virtual) and SUSY
pair decays are quite unlikely to be significant given LEP2 limits on the
masses of SUSY particles. However, by allowing Higgs (in particular,
pseudoscalar) masses such that one or more of the channels i)-vii) are
kinematically allowed we have found points for which discovery in modes 1)-9)
will not be possible \cite{wip}. Thus, a full ``no-lose'' theorem for NMSSM
Higgs boson discovery at the LHC will require exploring additional discovery
modes sensitive to those portions of parameter space for which Higgs decays to
other Higgs bosons are important, and might necessitate combining results from
both the ATLAS and CMS detectors and/or accumulating more integrated
luminosity.

\bigskip
\centerline{\bf Acknowledgments}
\medskip

We wish to acknowledge helpful discussions with M. Sapinski and D. Zeppenfeld.
C.H. would like to thank the late theory group of the Rutherford Appleton
Laboratory and the Theoretical Physics Department of Oxford, where part of this
work was achieved, for their kind hospitality. This work was supported in part
by the U.S. Department of Energy.

\newpage
\bigskip
\centerline{\bf Table Captions}
\medskip
\noi {\bf Table 1:} We tabulate the input bare model parameters, the
corresponding Higgs masses, and the corresponding Higgs couplings, relative to
SM Higgs boson coupling strength, for 6 bench mark points. Also given for the
CP-even $h_i$ are ratios of the $gg$ production rate and various branching
fractions relative to the values found for a SM Higgs of the same mass. For
the CP-odd $a_i$, ``gg Production Rate'' refers to the value relative to what
would be found if both the $b\anti b$ and the $t\anti t$ $\gamma_5$ couplings
had SM-like strength.
\medskip

\noi {\bf Table 2:} Scalar Higgs statistical significances, $N_{SD}=S/\sqrt B$,
in various channels for the 6 bench mark points. For each individual Higgs, we
give (in order): $N_{SD}$ for the channels 1) -- 9) described in the text;
Gaussian combined $N_{SD}$ for non-$WW$-fusion LHC channels; combined $N_{SD}$
for non-$WW$-fusion LHC channels plus LEP2; combined $N_{SD}$ for all LHC
channels, including the fusion channels $WW\to h\to \tau\anti\tau$ and $WW\to
h\to WW^{(*)}$ channels; and combined $N_{SD}$ for all LHC channels plus LEP2.
\medskip

\noi {\bf Table 3:} Summary for all Higgs bosons. The entries are: maximum
non-$WW$ fusion LHC $N_{SD}$; maximum LHC $WW$ fusion $N_{SD}$; best combined
$N_{SD}$ after summing over all non-$WW$-fusion LHC channels; and best
combined $N_{SD}$ after summing over all LHC channels. The Higgs boson for
which these best values are achieved is indicated in the parenthesis. One
should refer to the preceding table in order to find which channel(s) give the
best values.
\newpage

\begin{table}[p]
\center{\bf Table 1}\medskip
\begin{center}
\footnotesize
\begin{tabular} {|l|l|l|l|l|l|l|} 
\hline
Point Number & 1 & 2 & 3 & 4 & 5 & 6  \\
\hline \hline
Bare Parameters &\multicolumn{6}{c|}{} \\
\hline
$\lambda$            & 0.0340 & 0.0450 & 0.0230 & 0.0230 & 0.1330 & 0.0230 \\
\hline
$\kappa$             & 0.0198 & 0.0248 & 0.0129 & 0.0069 & 0.1459 & 0.0114 \\
\hline
$\tan\beta$          &   6.00 &   5.25 &   -5.5 &   5.75 &     -8 &     -6 \\
\hline
$\mu_{\rm eff} (GeV)$&    140 &   -110 &    115 &   -235 &    100 &    150 \\
\hline
$A_{\lambda} (GeV)$  &    -35 &     25 &    -95 &     40 &   -135 &   -100 \\
\hline
$A_{\kappa} (GeV)$   &   -150 &     70 &    -90 &     80 &    -75 &   -110 \\
\hline \hline
Scalar Masses and Couplings &\multicolumn{6}{c|}{} \\
\hline \hline
$m_{h_1}$ (GeV)      &    115 &    100 &    103 &    113 &    114 &    112 \\
\hline
$c_V $               &  -0.66 &   0.32 &  -0.34 &   0.67 &  -0.87 &  -0.71 \\
\hline
$c_t $               &  -0.65 &   0.30 &  -0.31 &   0.65 &  -0.81 &  -0.66 \\
\hline
$c_b $               &  -1.07 &   0.66 &  -1.27 &   1.16 &  -4.50 &  -2.40 \\
\hline
gg Production Rate   &   0.39 &   0.08 &   0.08 &   0.39 &   0.56 &   0.36 \\
\hline
$\br\gamma \gamma$   &   0.43 &   0.26 &   0.09 &   0.38 &   0.05 &   0.11 \\
\hline
$\br b\anti 
b=\br\tau\anti\tau$  &   1.12 &   1.08 &   1.10 &   1.12 &   1.18 &   1.15 \\
\hline
$\br WW^{(*)}$       &   0.42 &   0.25 &   0.08 &   0.37 &   0.04 &   0.10 \\
\hline \hline

$m_{h_2}$ (GeV)      &    125 &    114 &    114 &    126 &    144 &    122 \\
\hline
$c_V $               &  -0.74 &  -0.83 &   0.79 &  -0.73 &   0.46 &   0.59 \\
\hline
$c_t $               &  -0.72 &  -0.74 &   0.70 &  -0.71 &   0.57 &   0.54 \\
\hline
$c_b $               &  -1.49 &  -3.28 &   3.46 &  -1.47 &  -6.66 &   2.24 \\
\hline
gg Production Rate   &   0.46 &   0.44 &   0.40 &   0.45 &   1.18 &   0.23 \\
\hline
$\br\gamma \gamma$   &   0.33 &   0.08 &   0.07 &   0.34 &   0.01 &   0.10 \\
\hline
$\br b\anti 
b=\br\tau\anti\tau$  &   1.30 &   1.18 &   1.18 &   1.32 &   3.06 &   1.31 \\
\hline
$\br WW^{(*)}$       &   0.32 &   0.08 &   0.06 &   0.33 &   0.01 &   0.09 \\
\hline \hline

$m_{h_3}$ (GeV)      &    205 &    153 &    148 &   201 &     202 &    155 \\
\hline
$c_V $               &  -0.14 &  -0.46 &  -0.51 &  -0.15 &   0.18 &  -0.39 \\
\hline 
$c_t $               &  -0.30 &  -0.63 &  -0.67 &  -0.32 &   0.17 &  -0.55 \\
\hline
$c_b $               &   5.80 &   4.17 &   4.20 &   5.53 &   0.68 &   5.12 \\
\hline
gg Production Rate   &   0.31 &   0.84 &   0.95 &   0.33 &   0.02 &   0.80 \\
\hline
$\br\gamma \gamma$   &   0.13 &   0.05 &   0.05 &   0.15 &   0.98 &   0.03 \\
\hline
$\br b\anti 
b=\br\tau\anti\tau$  & 308.66 &   5.83 &   3.92 & 274.41 &  13.97 &   8.12 \\
\hline
$\br WW^{(*)}$       &   0.18 &   0.07 &   0.06 &   0.21 &   0.96 &   0.05 \\
\hline \hline

Pseudo-Scalar Masses and Couplings &\multicolumn{6}{c|}{} \\
\hline \hline
$m_{a_1}$ (GeV)      &    191 &    112 &    130 &    130 &    113 &    145 \\
\hline
$c_t $               &   0.03 &  -0.03 &  -0.10 &  -0.01 &  -0.10 &  -0.16 \\
\hline
$c_b $               &   1.16 &  -0.83 &  -2.95 &  -0.19 &  -6.55 &  -5.77 \\
\hline
gg Production Rate   &   0.00 &   0.00 &   0.03 &   0.00 &   0.31 &   0.08 \\
\hline \hline

$m_{a_2} $(GeV)      &    206 &    141 &    137 &    198 &    174 &    158 \\
\hline
$c_t $               &   0.16 &   0.19 &  -0.15 &   0.17 &  -0.07 &  -0.05 \\
\hline
$c_b $               &   5.89 &   5.18 &  -4.64 &   5.75 &  -4.59 &  -1.65 \\
\hline
gg Production Rate   &   0.02 &   0.07 &   0.06 &   0.02 &   0.03 &   0.00 \\
\hline \hline

Charged Higgs Mass   &\multicolumn{6}{c|}{} \\
\hline
$m_{c} $(GeV)        &    221 &    162 &    157 &    213 &    157 &    167 \\
\hline

\end{tabular}
\end{center}
\end{table}

\begin{table}[p]
\center{\bf Table 2}\medskip
\begin{center}
\footnotesize
\begin{tabular} {|l|l|l|l|l|l|l|} 
\hline
Point & 1 & 2 & 3 & 4 & 5 & 6   \\
\hline \hline
Channel & \multicolumn{6}{c|}{$h_1$ Higgs boson} \\
\hline
\nsd1  &  3.74 &  0.35 &  0.13 &  3.18 &  0.62 &  0.83 \\
\nsd2  &  4.37 &  0.59 &  0.22 &  3.92 &  0.85 &  1.22 \\
\nsd3  &  2.79 &  0.85 &  0.85 &  3.03 &  4.83 &  3.30 \\
\nsd4  &  0.08 &  0.07 &  0.76 &  0.09 &  4.52 &  0.40 \\
\nsd5  &  0.83 &  0.00 &  0.00 &  0.64 &  0.12 &  0.16 \\
\nsd6  &  1.10 &  0.09 &  0.03 &  0.90 &  0.16 &  0.22 \\
\nsd7  &  0.00 &  3.37 &  3.40 &  3.29 &  0.00 &  4.79 \\
\nsd8  &  9.29 &  1.22 &  1.59 &  8.93 & 16.78 & 10.08 \\
\nsd9  &  2.39 &  0.00 &  0.00 &  1.74 &  0.41 &  0.49 \\
$\sqrt{\sum_{i=1}^6 [N_{SD}(i)]^2}$ 
       &  6.54 &  1.09 &  1.17 &  5.99 &  6.69 &  3.65 \\
$\sqrt{\sum_{i=1}^7 [N_{SD}(i)]^2}$ 
       &  6.54 &  3.55 &  3.59 &  6.84 &  6.69 &  6.02 \\
$\sqrt{\sum_{i=1-6,8,9} [N_{SD}(i)]^2}$ 
       & 11.61 &  1.64 &  1.97 & 10.89 & 18.07 & 10.73 \\
$\sqrt{\sum_{i=1}^9 [N_{SD}(i)]^2}$ 
       & 11.61 &  3.75 &  3.93 & 11.38 & 18.07 & 11.75 \\
\hline \hline
Channel & \multicolumn{6}{c|}{$h_2$ Higgs boson} \\
\hline
\nsd1  &  3.69 &  0.83 &  0.61 &  3.62 &  0.22 &  0.55 \\
\nsd2  &  4.01 &  1.25 &  0.92 &  3.93 &  0.05 &  0.74 \\
\nsd3  &  2.49 &  3.95 &  3.58 &  2.30 &  0.99 &  1.77 \\
\nsd4  &  0.16 &  2.76 &  2.93 &  0.16 &  3.62 &  2.99 \\
\nsd5  &  1.84 &  0.16 &  0.11 &  1.94 &  0.56 &  0.20 \\
\nsd6  &  1.44 &  0.22 &  0.16 &  1.46 &  0.38 &  0.18 \\
\nsd7  &  0.00 &  0.00 &  3.31 &  0.00 &  0.00 &  0.00 \\
\nsd8  & 15.39 & 15.17 & 13.46 & 15.05 &  7.41 &  9.89 \\
\nsd9  &  5.79 &  0.63 &  0.44 &  6.05 &  0.19 &  0.82 \\
$\sqrt{\sum_{i=1}^6 [N_{SD}(i)]^2}$ 
       &  6.44 &  5.05 &  4.76 &  6.31 &  3.82 &  3.61 \\
$\sqrt{\sum_{i=1}^7 [N_{SD}(i)]^2}$ 
       &  6.44 &  5.05 &  5.80 &  6.31 &  3.82 &  3.61 \\
$\sqrt{\sum_{i=1-6,8,9} [N_{SD}(i)]^2}$ 
       & 17.65 & 16.00 & 14.28 & 17.40 &  8.34 & 10.56 \\
$\sqrt{\sum_{i=1}^9 [N_{SD}(i)]^2}$ 
       & 17.65 & 16.00 & 14.66 & 17.40 &  8.34 & 10.56 \\
\hline \hline
Channel  & \multicolumn{6}{c|}{$h_3$ Higgs boson} \\
\hline
\nsd1  &  0.00 &  0.59 &  0.66 &  0.01 & 0.00 &  0.32 \\
\nsd2  &  0.00 &  0.21 &  0.25 &  0.00 & 0.00 &  0.08 \\
\nsd3  &  0.00 &  0.00 &  1.13 &  0.00 & 0.00 &  0.00 \\
\nsd4  &  3.79 &  3.43 &  3.62 &  3.56 & 1.55 &  4.86 \\
\nsd5  &  3.65 &  2.51 &  2.07 &  4.46 & 1.54 &  1.66 \\
\nsd6  &  0.80 &  2.13 &  1.52 &  1.17 & 0.38 &  1.55 \\
\nsd7  &  0.00 &  0.00 &  0.00 &  0.00 & 0.00 &  0.00 \\
\nsd8  &  0.00 &  0.00 &  9.06 &  0.00 & 0.00 &  0.00 \\
\nsd9  &  0.00 &  0.77 &  0.79 &  0.00 & 0.00 &  0.43 \\
$\sqrt{\sum_{i=1}^6 [N_{SD}(i)]^2}$ 
       &  5.32 &  4.80 &  4.64 &  5.83 & 4.76 &  5.37 \\
$\sqrt{\sum_{i=1}^7 [N_{SD}(i)]^2}$ 
       &  5.32 &  4.80 &  4.64 &  5.83 & 4.76 &  5.37 \\
$\sqrt{\sum_{i=1-6,8,9} [N_{SD}(i)]^2}$ 
       &  5.32 &  4.86 & 10.21 &  5.83 & 4.76 &  5.39 \\
$\sqrt{\sum_{i=1}^9 [N_{SD}(i)]^2}$ 
       &  5.32 &  4.86 & 10.21 &  5.83 & 4.76 &  5.39 \\
\hline
\end{tabular} 
\end{center}
\end{table}
\clearpage

\begin{table}
\center{\bf Table 3}\medskip
\begin{center}
\footnotesize
\hspace*{-.5cm}
\begin{tabular} {|l|l|l|l|l|l|l|}
\hline
Point Number & 1 & 2 & 3 & 4 & 5 & 6   \\
\hline
Best non-$WW$ fusion $N_{SD}$ 
&  4.37 ($h_1$) &  3.95 ($h_2$) &  3.62 ($h_3$) &  4.46 ($h_3$) &  4.83 ($h_1$)  &  4.86 ($h_3$) \\
\hline Best $WW$ fusion $N_{SD}$ 
& 15.39 ($h_2$) & 15.17 ($h_2$) & 13.46 ($h_2$) & 15.05 ($h_2$) & 16.78 ($h_1$)  & 10.08 ($h_1$) \\
\hline \begin{minipage}{3.8cm}{\baselineskip=0pt Best combined $N_{SD}$
w.o.\\ $WW$-fusion modes\vspace*{.15cm}}\end{minipage} 
&  6.54 ($h_1$) &  5.05 ($h_2$) &  4.76 ($h_2$) &  6.31 ($h_2$) &  6.69 ($h_1$)  &  5.37 ($h_3$) \\
\hline \begin{minipage}{3.8cm}{\baselineskip=0pt Best combined $N_{SD}$
with\\ $WW$-fusion modes\vspace*{.15cm}}\end{minipage}
& 17.65 ($h_2$) & 16.00 ($h_2$) & 14.28 ($h_2$) & 17.40 ($h_2$) & 18.07 ($h_1$)  & 10.73 ($h_1$) \\
\hline
\end{tabular}
\end{center}
\end{table}

\end{document}